\documentclass{aastex631}

\makeatletter
\def\blfootnote{\xdef\@thefnmark{}\@footnotetext}
\makeatother

\received{February 28, 2022}
\revised{April 18, 2022}
\accepted{April 22, 2022}
\submitjournal{ApJ}

\shorttitle{Sample article}
\shortauthors{Messias et al.}

\begin{document}

\title{A dearth of close-in planets around rapidly rotating stars or a dearth of data? }

\correspondingauthor{Yuri S. Messias}
\email{yuri.messias.016@ufrn.edu.br}

\author[0000-0002-2425-801X]{Y. S. Messias}
\affiliation{Departamento de F\'isica Te\'orica e Experimental, Universidade Federal do Rio Grande do Norte, Campus Universit\'ario, Natal, RN, 59072-970, Brazil}
\author[0000-0002-2068-0205]{L. L. A. de Oliveira}
\affiliation{Departamento de F\'isica Te\'orica e Experimental, Universidade Federal do Rio Grande do Norte, Campus Universit\'ario, Natal, RN, 59072-970, Brazil}
\author[0000-0002-2023-7641]{R. L. Gomes}
\affiliation{Departamento de F\'isica Te\'orica e Experimental, Universidade Federal do Rio Grande do Norte, Campus Universit\'ario, Natal, RN, 59072-970, Brazil}
\author[0000-0001-7114-8975]{M. I. Arruda Gonçalves}
\affiliation{Departamento de F\'isica Te\'orica e Experimental, Universidade Federal do Rio Grande do Norte, Campus Universit\'ario, Natal, RN, 59072-970, Brazil}
\author[0000-0001-5578-7400]{B. L. Canto Martins}
\affiliation{Departamento de F\'isica Te\'orica e Experimental, Universidade Federal do Rio Grande do Norte, Campus Universit\'ario, Natal, RN, 59072-970, Brazil}
\author[0000-0001-5845-947X]{I. C. Le\~ao}
\affiliation{Departamento de F\'isica Te\'orica e Experimental, Universidade Federal do Rio Grande do Norte, Campus Universit\'ario, Natal, RN, 59072-970, Brazil}
\author[0000-0001-8218-1586]{J. R. De Medeiros}
\affiliation{Departamento de F\'isica Te\'orica e Experimental, Universidade Federal do Rio Grande do Norte, Campus Universit\'ario, Natal, RN, 59072-970, Brazil}

\begin{abstract} 
A dearth of close-in planets orbiting rapid rotators was reported almost a decade ago. According to this view only slowly spinning stars with rotation periods longer than 5-10 days would host planets with orbital periods shorter than 2 or 3 days. This Letter brings an enlarged and more detailed analysis that led us to the question: Is there really a dearth in that distribution or is it a dearth of data? For this new analysis, we combined different samples of Kepler and TESS stars with confirmed planets or planet candidates with measured stellar rotation periods, using Gaia data to perform an in-depth selection of 1013 planet-hosting main-sequence stars. With the newer, enlarged, and more refined data, the reported dearth of close-in planets orbiting rapid rotators tends to disappear, thus suggesting that it may reflect a scarcity of data in the prior analysis. A two sample statistical test strongly supports our results, showing that the distribution of close-in planets orbiting rapid rotators is almost indistinguishable from that for close-in planets orbiting slow rotators.
\end{abstract}


\section{Introduction} \label{sec:intro}

Theoretical considerations and observational evidence show that stars may interact with their planets through gravitation, radiation, and magnetic fields. For main-sequence late-type stars with close-in planets, effects on the host star owing to tidal and magnetic interaction with the exoplanet are now supported by different studies. In this context, \cite{2009MNRAS.396.1789P} has pointed out that stars with transiting hot Jupiters present an excess of rapid rotators in comparison to stars without close-in planets, whereas~\cite{2009A&A...505..339L,2010A&A...512A..77L} has suggested that, among these stars, synchronization tends to increase with increasing effective temperature. Knowledge of stellar rotation periods can also offer an important constraint for our understanding of the behavior of the stellar angular momentum \citep{2019PASP..131a4401G}, a stellar parameter that appears to be in deficit in stars without detected planets in relation to stars with detected planets~\citep{2010MNRAS.408.1770A}.

 \cite{2013ApJ...775L..11M} have reported a possible dearth of close-in planets around rapidly rotating stars, based on a sample of 737 Kepler Objects of Interest (KOIs) assumed by the authors to be planet-hosting main-sequence stars, and for which the authors were able to measure the rotation period. According to those authors, only slowly rotating stars, with rotation periods longer than 5–10 days, host planets on orbits shorter than 2-3 days. The root cause that led to this trend is not yet well established, although tidal interaction is suggested to be the most prominent hypothesis \citep[e.g.,][]{2013MNRAS.436.1883W,2014ApJ...786..139T}. Another scenario is proposed by~\cite{2014MNRAS.443.1451L}, who interpret the phenomenon observed by \cite{2013ApJ...775L..11M} on the basis of secular perturbations in multiplanet systems located in nonresonant orbits.

Using the photometric observations carried out by the Transiting Exoplanet Survey Satellite (TESS; \citealt{2015JATIS...1a4003R}), \cite{2020ApJS..250...20C} measured the rotation period for dozens of TESS Objects of Interest (TOIs), namely stars with planets or planet candidates with orbital periods revealed by TESS. The present study presents a detailed analysis of the origin of the dearth of close-in planets around fast-rotating main-sequence stars suggested by \cite{2013ApJ...775L..11M}, taking into account an enlarged stellar sample and a careful analysis of the luminosity class and potential binary status of the considered working sample. This latter aspect, in particular,  is mandatory for our  understanding of the underscored dearth of close-in planets around fast rotators because  significant contamination by evolved and binary stars can affect the variability properties of dwarf stars \citep{2011AJ....141..108C,2012ApJ...753...90M}. This study is organized as follows. Section \ref{observation} presents the stellar sample and observational data set used in the analysis. Section \ref{sec:results} provides the main results, with a summary presented in Section \ref{sec:summary}.

\section{Working Sample} \label{observation}

For the purposes of the present study, we have initially aimed to enlarge the working stellar sample as much as possible by searching for additional KOI stars, as well as TOI stars, with rotation periods available in the literature. 
In this sense, we have added 447 KOIs from \cite{2013MNRAS.436.1883W} and 339 KOIs from \cite{2015ApJ...801....3M} to the list of 737 KOIs used by \cite{2013ApJ...775L..11M}, amounting to a sample of 1523 KOI stars. Those two additional samples, with available measurements of rotation period along with likely transit orbital period, comprise a subset of stars that are not in common with those studied by \cite{2013ApJ...775L..11M}.
\cite{2013MNRAS.436.1883W} determined rotation periods using the Lomb--Scargle method \citep{1976Ap&SS..39..447L,1982ApJ...263..835S}, whereas \cite{2015ApJ...801....3M} used the autocorrelation function (ACF) method -- described in detail by \citet{2014ApJS..211...24M} -- in a similar manner to \cite{2013ApJ...775L..11M}, however with newer data and a more refined procedure. 

Then, we have  added to the referred list of KOIs a sample of 222 TOI stars with rotation periods given by \cite{2020ApJS..250...20C}, thus composing a combined list of 1745 KOI and TOI stars with available stellar rotation and potential planetary orbital periods. The rotation periods for the TOI stars were computed based on a manifold analysis that used visual inspection of the light curves together with fast-Fourier transform, Lomb--Scargle periodograms, and wavelet maps.
 Next, we checked for false-positive planet candidates based on flags available for KOIs and TOIs in the NASA Exoplanet Archive (NEA, \citealt{2013PASP..125..989A}). After removing those false positives, as of 2021 December 10, as well as stars with no Gaia information and anomalous parallaxes, the combined stellar sample amounts to 1194 KOIs and TOIs.
 
 With this list of 1194 KOI and TOI stars in hand, we have applied a well-established procedure based on the analysis of the location of stars in the Gaia color-magnitude diagram (CMD) to separate the main-sequence from evolved stars, as well as to identify potential binary systems \citep[e.g.,][]{2017ApJ...835...16D,2019Tese..Luciano,2021ApJ...913...70G}. Such step is mandatory, in particular, because \cite{2013ApJ...775L..11M} have defined their stellar sample as being composed of planet-host main-sequence stars, but different studies have shown that the enlarged stellar sample of \cite{2014ApJS..211...24M}, which includes the sample of \cite{2013ApJ...775L..11M}, is contaminated  by subgiants, giants, and binary stars \citep[e.g.,][]{2017ApJ...835...16D,2019Tese..Luciano,2021ApJ...913...70G}. Indeed, several studies have demonstrated that contamination by evolved or binary stars can affect variability properties of dwarf stars and, by consequence, the statistics of stellar and planetary parameters \citep{2011AJ....141..108C,2012ApJ...753...90M,2015ApJ...805...16C,2021AJ....161..231W,2021AJ....162..192Z}. Accordingly, to properly understand the nature of the rotation period versus orbital period distributions of dwarf stars, an in-depth selection of main-sequence stars must be considered in the analysis.

\subsection{Selecting main sequence stars} \label{Main-sequence stars}

\begin{figure} 

  \begin{center}
    \includegraphics[width=0.55\textwidth]{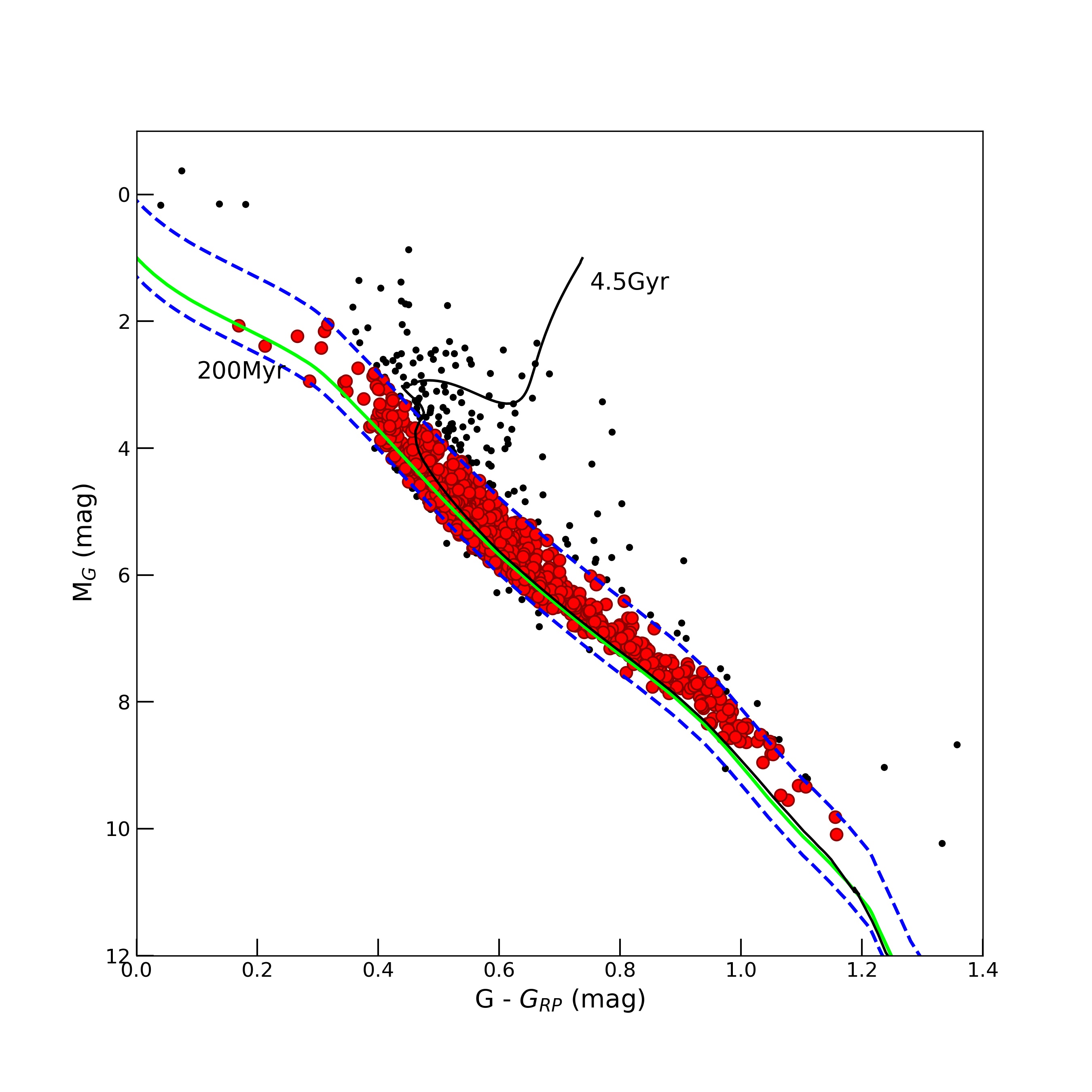} 
   \vspace{-39pt}
  \end{center}
  \caption{Gaia color-magnitude diagram for the combined sample of~1013~KOIs and TOIs selected as main-sequence stars, that met the Gaia photometric and parallax criteria described in the text. For reference, we show an MIST isochrone \citep{2011ApJS..192....3P,2013ApJS..208....4P,2015ApJS..220...15P,2016ApJ...823..102C,2016ApJS..222....8D} with an age of~200~Myr and a metallicity of [Fe/H]~=~$+0.25$ to identify the likely main sequence of 200~Myr (green curve). An MIST isochrone with an age of~4.5~Gyr is also plotted to illustrate the subgiant and giant branches (solid black curve). Main-sequence stars are selected as those between the~200~Myr isochrone shifted down and up (dashed blue curves) by~0.3 and~0.9~mag, respectively.}
  \label{figura:fig1}
\end{figure}

Following the same approach by \cite{2021ApJ...913...70G} for the construction of the Gaia CMD, we begin by making the selection of Gaia magnitudes and color indexes based on the quality of the photometric solutions \citep{2018A&A...616A...1G}.
 As in \cite{2021ApJ...913...70G}, we only selected stars with $\sigma(G)/G < 0.01$ and $\sigma(G_{\rm RP})/G_{\rm RP} < 0.01$,
where $G$ and $G_{\rm RP}$ refer to the passbands used in the Gaia first and second data releases (DR1 and DR2, respectively; \citealt{2016A&A...595A...2G,2018A&A...616A...1G}). Following these conditions, we have placed our combined list of~1194~KOI and~TOI~stars in the Gaia CMD displayed in~Fig.~\ref{figura:fig1}, which also shows an isochrone obtained from the Modules for Experiments in Stellar Astrophysics Isochrones \& Stellar Tracks (MIST) \citep{2011ApJS..192....3P,2013ApJS..208....4P,2015ApJS..220...15P,2016ApJ...823..102C,2016ApJS..222....8D} with an age of~200~Myr and a metallicity of~[Fe/H]~=~$+0.25$, which adequately matches the trend of the main sequence. Single main-sequence stars were selected within~0.3~mag below and~0.9~mag above the isochrone of 200 Myr, as shown in~Fig.~\ref{figura:fig1}. As underlined by~\cite{2021ApJ...913...70G}, such a
range of magnitude should enclose stars of different ages and metallicities, avoiding, at the same time, contamination from the subgiant and giant branches, as well as from binary stars.
Indeed, different studies have pointed out that, beyond the subgiant and giant contamination, Gaia CMDs also reveal a secondary population of stars 
above the main sequence, which arises due to unresolved equal - or nearly equal - mass \citep[e.g.,][]{2018AJ....156..145A,2018ApJ...868..151D,2021ApJ...913...70G}. Once applied the above conditions, our final sample consists of~1013~ likely main-sequence stars, meeting our Gaia photometry criteria. This sample of combined TOI and KOI stars is given in Table~\ref{tab1}.

\begin{table}[!htbp]  
	\centering 
	\caption{\label{tab1} The list of 1013 selected TOI and KOI main sequence stars \blfootnote{Notes. The following information is listed: the TIC and KIC ID, rotation period ($P_{rot}$) and planetary orbital period ($P_{orb}$).\\ References: $P_{orb}$ for TOI stars from NASA Exoplanet Archive -- TESS Project Candidates \citep{https://doi.org/10.26134/exofop3}, and for KOI stars from NASA Exoplanet Archive -- KOIs \citep{cumulative}}.}
	\begin{tabular}{ccc}
		\hline \hline
		KIC / TIC	&	$P_{rot}$	&	$P_{orb}$ 	\\	
     	          &	 (days)	    &	 (days)    \\	\hline
     	 &   Stars from \cite{2013ApJ...775L..11M}\\ \hline
 KIC 757450	 &	19.38		&	8.89		\\	
 KIC 2142522	 &	10.38		&	13.32     	\\	
 KIC 2161536	 &	27.27		&	16.86	   \\	
 KIC 2165002 &	22.75		&	16.57	 	\\	
 KIC 2302548 &	12.37		&	10.38	 	\\	
 KIC 2438513 &	13.54		&	12.18	     \\
     ...     &   ...     	&   ...			  \\
\hline						
	\end{tabular}
\end{table}

\section{Results} \label{sec:results}

\begin{figure}[htp!]
 \centering
\includegraphics[width=0.75\textwidth]{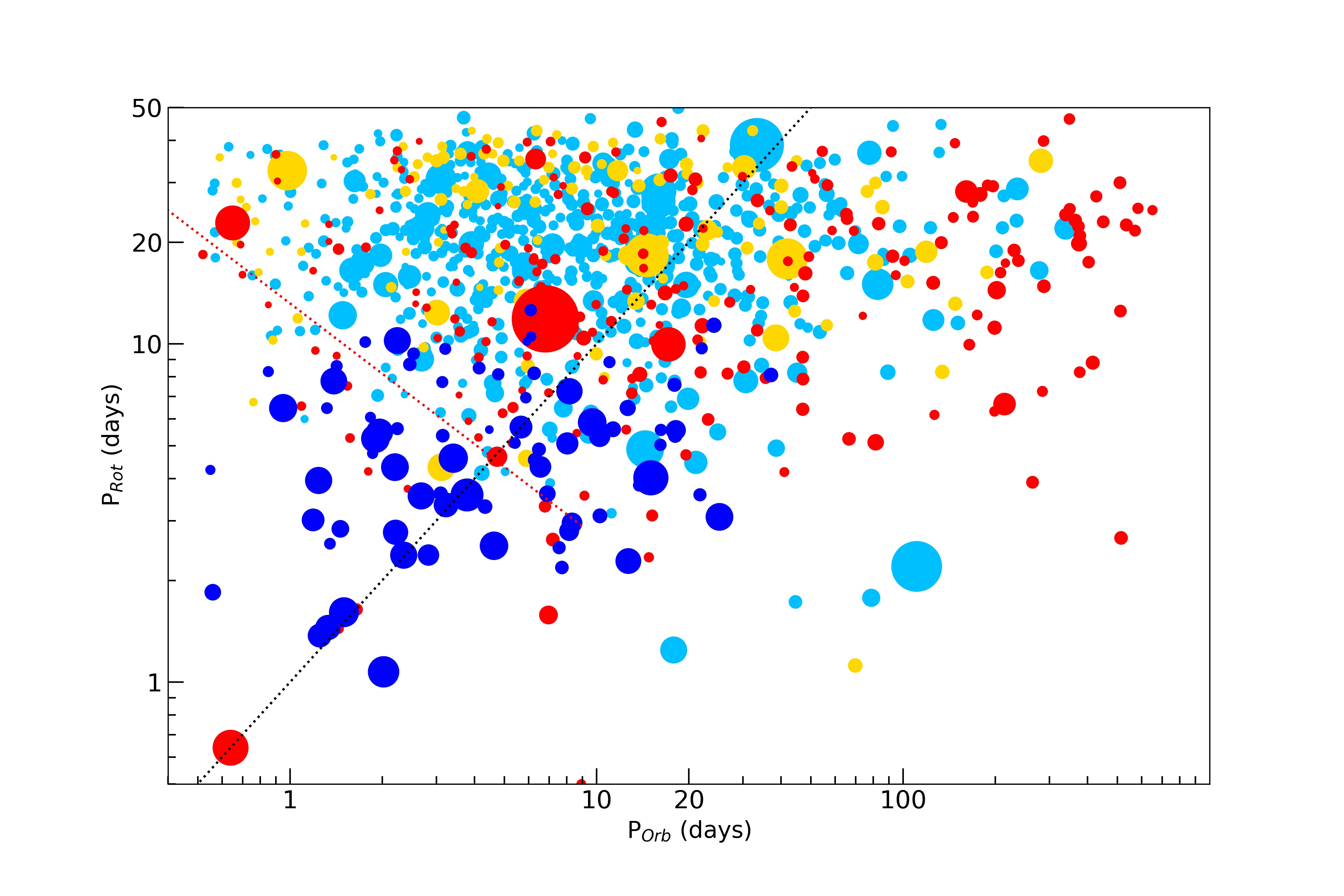} 
\caption{Stellar rotation period, $P_{\rm rot}$, as a function of planetary orbital period, $P_{\rm orb}$, of the innermost planet for the combined sample of~1013~KOIs~and~TOIs, selected as main-sequence stars, with rotation periods available in the literature. Light blue, yellow, red, and dark blue circles represent the stars from \cite{2013ApJ...775L..11M}, \cite{2013MNRAS.436.1883W}, \cite{2015ApJ...801....3M}, and \cite{2020ApJS..250...20C}, respectively. Circle sizes are proportional to planet radius squared.
The red dashed line is the fit with the lower envelope of points described by \cite{2013ApJ...775L..11M}.
The black dotted line marks the 1:1 synchronization rate between $P_{\rm orb}$ and $P_{\rm rot}$.}
\label{figura:fig2}
\end{figure}

 Based on the refined stellar sample obtained in the present study, we display in Fig.~\ref{figura:fig2} the distribution of stellar rotation period, $P_{\rm rot}$, versus planetary orbital period, $P_{\rm orb}$,
following closely the same format as in~\cite{2013ApJ...775L..11M}. The new scenario emerging from this enlarged sample points to the trend for the disappearance of the dearth of planets at short orbital periods around fast rotators previously suggested by \cite{2013ApJ...775L..11M}. Let us recall that, according to these authors, only slow stellar rotators, with periods larger than 5--10 days, would have planets with periods shorter than~2~or~3~days.

A careful inspection of Fig.~\ref{figura:fig2} shows that the region where the dearth would be located now tends to be populated as a result of the new stellar sample, considering only likely main-sequence stars. 
To evaluate the significance of this strong contrast between the present results and those by ~\cite{2013ApJ...775L..11M}, we also performed a two-sample test using the Anderson--Darling (A-D) test \citep{1987JASA...82p918}, and then the Kolmogorov--Smirnov (K-S) test \citep{1992nrca.book.....P}, which calculate the probability that two distributions are derived from the same parent distribution. First, we have analyzed the orbital period distributions for $P_{\rm orb} \leq 10$~days, split at $P_{\rm rot} = 18.95$~days, considering TOI and KOI stars, following the same definitions used ~\cite{2013ApJ...775L..11M}. This strategy provided an equal number of 292 stars in each sample. From the A-D test, the \textit{p-value} (probability for the null hypothesis that both samples come from the same distribution) is $   \geq$ 0.250, in contrast to the value of 0.018 found by ~\cite{2013ApJ...775L..11M}, whereas the K-S test gives a probability of 0.978, consistent with the two distributions being drawn from the same parent distribution. 
Second, to avoid any concern about the contribution of KOI and TOI samples to the disappearance of the referred dearth, we have also  carried out the same two-sample test but considering only the present sample of 934 KOI stars. Such a step is mandatory due to the TESS observational strategy for measuring especially fast rotation \citep[e.g.,][]{2020ApJS..250...20C, 2021AJ....162..147H}, and for detecting planets with short-orbital periods (\citealt{2015JATIS...1a4003R,2021ApJS..254...39G}). Following the same above strategy, we obtained two samples with an equal number of 262 stars in each sample, for $P_{\rm orb} \leq 10$~days, split at $P_{\rm rot} = 20.21$~days. This additional analysis gives an A-D \textit{p-value} $\geq$0.25, and a K-S probability of 0.431, showing that the disappearance of the dearth seems to result, effectively, from the increase of the stellar sample considered, independently of whether the increasing comes from TOI or KOI stars. Therefore, the lower-left region of the Fig.~\ref{figura:fig2}, which was suggested to reveal a dearth of close-in planets for $P_{\rm orb} \leq 10$~days, actually displays a distribution of stars with close-in planets indistinguishable from that for the upper-left region also for $P_{\rm orb} \leq 10$~days.

\section{Summary} \label{sec:summary}
We have found solid observational evidence that the possible dearth of close-in planets around fast-rotating main-sequence stars suggested by \cite{2013ApJ...775L..11M}, seems to reflect primarily a dearth of observational data by considering the stellar sample used by those authors. Thanks to an enlarged sample of ~1013~ likely main-sequence stars, combining KOIs and TOIs, analyzed rigorously to eliminate contamination by evolved and binary stars, the present study shows a clear trend for the disappearance of the referred dearth scenario. Both the A-D and K-S two-sample tests support our observational finding, showing that the distribution of close-in planets orbiting rapid rotators is almost indistinguishable from that for close-in planets orbiting slowly spinning stars.

In the light of the new scenario pointing to the apparent nonexistence of a dearth of planets at short orbital periods around fast-rotating stars, a few key aspects need to be underscored. It seems clear that the disappearance of the dearth arises independently of the TOI sample. The two-sample test performed shows unambiguously that the disappearance of the dearth already manifests itself from a scenario in which only new KOI data are added to the sample used by \cite{2013ApJ...775L..11M}. Even so, the presence of the TOI stars in the present sample, reinforces the primary feature emerging from our results: the existence of planets with short-orbital periods, typically $P_{orb}$ shorter than about 4-5 days, orbiting fast-rotating main-sequence stars, that is, stars with $P_{rot}$ shorter than about 5 days. Nevertheless, such a feature, in principle, should not be interpreted as revealing a uniform existence of short-orbital planets around stars with any value of rotation period because the trend referred to may only indicate that due to its observational strategy, TESS is revealing many short-period planets than observed to date. Additional TESS observations, in particular from the extended mission that runs through this coming September, will provide measures of planetary and stellar rotation periods for a plethora of TOIs. We expect that these new data will reveal particular trends in the frequency of short-orbital planets around fast-rotating stars and on its dependence on different stellar rotation and planetary orbital periods. 

 Finally, it is noteworthy that the primary finding of our study may represent an additional challenge for the present-day theoretical models dealing with the evolution of star–-planet systems \citep[e.g.:][]{2014MNRAS.443.1451L,2021A&A...650A.126A}. Different studies point to a theoretical distribution in stellar rotation versus planetary periods, according to which there would be a region at low stellar rotation periods and low planetary orbital periods not populated by star–-planet systems. Such a perspective is, in principle, in contrast to the observational evidence emerging from the present study.

\begin{acknowledgments}
\textit{Acknowledgments:} We warmly thank our families for showing us care, patience, and tenderness during the home office tasks for the preparation of this study in the face of the difficult moments presented by COVID-19. Research activities of the board of observational astronomy at the Federal University of Rio Grande do Norte are supported by continuous grants from the Brazilian funding agencies CNPq and FAPERN. This study was financed in part by the Coordena\c{c}\~ao de Aperfei\c{c}oamento de Pessoal de N\'ivel Superior - Brasil (CAPES) - Finance Code 001. Y.S.M., R.G.L. and M.I.A.G. acknowledge CAPES graduate fellowships. L.L.A.O. acknowledges a CAPES/PNPD fellowship. B.L.C.M., I.C.L., and J.R.M. acknowledge CNPq research fellowships. This Letter includes data collected by the TESS and Kepler missions. This research has made use of the NASA Exoplanet Archive, which is operated by the California Institute of Technology, under contract with the National Aeronautics and Space Administration under the Exoplanet Exploration Program. We acknowledge the use of public TIC Release data from pipelines at the TESS Science Office and at the TESS Science Processing Operations Center. TESS data were obtained from the Mikulski Archive for Space Telescopes (MAST). Funding for the TESS mission is provided by NASA's Science Mission directorate. The authors thank the anonymous referee for helpful comments and suggestions, which largely improved this study.

\end{acknowledgments}

\bibliographystyle{aasjournal}
\bibliography{References}

\end{document}